\documentclass[12pt]{article}

\usepackage{epsfig}
\usepackage{latexsym}

\def\balpha{\mbox{\boldmath $\alpha$}}
\def\bbeta{\mbox{\boldmath $\beta$}}

\begin{document}

\begin{titlepage}

\baselineskip 24pt

\begin{center}

{\Large {\bf New Angle on the Strong CP and Chiral Symmetry Problems 
from a Rotating Mass Matrix}}

\vspace{.5cm}

\baselineskip 14pt

{\large Jos\'e BORDES}\\
jose.m.bordes\,@\,uv.es\\
{\it Departament Fisica Teorica, Universitat de Valencia,\\
  calle Dr. Moliner 50, E-46100 Burjassot (Valencia), Spain}\\
\vspace{.2cm}
{\large CHAN Hong-Mo}\\
chanhm\,@\,v2.rl.ac.uk \\
{\it Rutherford Appleton Laboratory,\\
  Chilton, Didcot, Oxon, OX11 0QX, United Kingdom}\\
\vspace{.2cm}
{\large TSOU Sheung Tsun}\\
tsou\,@\,maths.ox.ac.uk\\
{\it Mathematical Institute, University of Oxford,\\
 24-29 St. Giles', Oxford, OX1 3LB, United Kingdom}

\end{center}

\vspace{.3cm}

\begin{abstract}

It is shown that when the mass matrix changes in orientation (rotates) 
in generation space for changing energy scale, then the masses of the 
lower generations are not given just by its eigenvalues.  In particular, 
these masses need not be zero even when the eigenvalues are zero.  In 
that case, the strong CP problem can be avoided by removing the unwanted 
$\theta$ term by a chiral transformation in no contradiction with the 
nonvanishing quark masses experimentally observed.  Similarly, a rotating 
mass matrix may shed new light on the problem of chiral symmetry breaking. 
That the fermion mass matrix may so rotate with scale has been suggested 
before as a possible explanation for up-down fermion mixing and fermion 
mass hierarchy, giving results in good agreement with experiment.

\end{abstract}

\end{titlepage}

\clearpage

\baselineskip 14pt

A fermion mass matrix extracted directly from a Yukawa term in the 
action takes usually the following form:
\begin{equation}
m \frac{1}{2}(1 + \gamma_5) + m^{\dagger} \frac{1}{2}(1 - \gamma_5).
\label{massmat1}
\end{equation}
However, by appropriately relabelling the right-handed fields in the
singlet representation, which will not change the physics, the mass
matrix can always be recast into a hermitian form independent of 
$\gamma_5$ \cite{Weinberga}.  It is this hermitian form of the fermion 
mass matrix that will exclusively be used in the present paper, which 
is henceforth taken as understood.  Furthermore, to be specific, the 
analysis will be carried out explicitly only for the realistic case 
of 3 fermion generations, although it can readily be extended with 
only minor modifications to other numbers of fermion generations.

By a rotating fermion mass matrix then, we mean a fermion mass matrix 
which changes its orientation in generation space as the scale changes.  
Explicitly,
\begin{equation}
m(\mu) = U(\mu, \mu_0) m(\mu_0) U^{-1}(\mu, \mu_0),
\label{rotation}
\end{equation}
where $U(\mu, \mu_0)$ is unitary, whose explicit form will depend on
the theory under consideration, but which we can leave unspecified for 
the present general discussion

That the mass matrix $m$ can so rotate is expected.  In much the same 
way as the familiar running of coupling strengths and mass values 
results as a consequence of renormalization, so will generally the 
rotation of the mass matrix.  Indeed, even in the Standard Model, so 
long as the CKM or MNS mixing matrix is not diagonal, the fermion mass 
matrix will rotate with changing scale \cite{Ramond}, although the 
rotation there is so slow as to be negligible for most practical 
purposes.  Once beyond the Standard Model framework, however, it will 
not be difficult to imagine situations where rotation becomes more 
appreciable. 

For the moment, we shall not address the question what dynamics will
generate appreciable rotation or whether such dynamics is realistic,
but concentrate first on the theoretical question of what physical
consequences will result from a rotating fermion mass matrix whatever 
its origin, of which consequences, as we shall see, there are some of 
considerable interest.  Only at the end of the paper shall we return
to summarize some evidences for mass matrix rotation, both empirical 
and theoretical. 

Not surprisingly, a rotating mass matrix will force on us some changes
in notions we have grown used to in the situation when the mass matrix 
does not rotate.  Indeed, one soon learns from experience when working 
with a rotating mass matrix that it would be unwise to take for granted 
any of these notions, no matter how familiar, without first checking
whether it can be extended unchanged to the rotating case.  One example
of particular interest to the present discussion is the statement often
made, based on experience gained from the nonrotating mass matrix, that 
chiral invariant interactions cannot generate nonzero physical masses 
from an initially chiral invariant mass matrix.  We shall immediately 
see below that such a statement cannot in general be maintained without 
modifications for a rotating mass matrix.  

Of course, whether the mass matrix rotates or not, any chiral invariant
interactions will leave an initially chiral invariant mass matrix still
chiral invariant.  More precisely, one means that starting with a mass
matrix of a certain rank, then under interactions of the same rank, the
rank of the mass matrix will be maintained.  For example, starting with
a rank 1 mass matrix in 3 generations, the renormalized mass matrix will
remain of rank 1, i.e.
\begin{equation}
m(\mu) = m_T(\mu) |\balpha(\mu) \rangle \langle \balpha(\mu)|,
\label{mfact}
\end{equation}
or that it still has 2 zero eigenvalues.  For a nonrotating mass matrix,
i.e. when $|\balpha(\mu) \rangle$ does not depend on $\mu$, it then 
follows that 2 of the physical particles must still have zero mass, 
since the masses are just given by the eigenvalues.  In case the mass 
matrix rotates, i.e. when $|\balpha(\mu) \rangle$ does indeed depend on 
$\mu$, however, this does {\it not} follow, since the masses of the 
physical particles are {\it not} all given just by the eigenvalues.

That this is the case may seem surprising at first sight, but it can be 
verified immediately as follows.  To be specific, let us consider the 
charged leptons, assuming that the mass matrix rotates but remains of 
rank 1 at all scales, i.e. of the form (\ref{mfact}).  To identify the 
masses of the physical states, we need first to specify these physical 
states.  The heaviest physical state, say ${\bf v}_\tau$, is easy; it 
is the single massive eigenstate $\balpha(\mu)$ of the mass matrix 
(\ref{mfact}) taken at the scale equal to its mass $\mu = m_\tau$, i.e. 
${\bf v}_\tau = \balpha(m_\tau)$.  The other physical states, ${\bf v}_\mu$
and ${\bf v}_e$ have then to be orthogonal to ${\bf v}_\tau$ and to each 
other, for $\tau, \mu, e$ are supposedly independent quantum states.  
Otherwise, if, say, ${\bf v}_\tau$ is allowed to have a nonzero component 
in ${\bf v}_\mu$ or ${\bf v}_e$, then $\tau$ can decay readily into 
$\mu \gamma$ or $e \gamma$ leading to blatant flavour-violations unseen 
in experiment.  Hence, ${\bf v}_\mu$ and ${\bf v}_e$ must have eigenvalue 
zero at the scale $\mu = m_\tau$.  But this zero eigenvalue at the scale 
$\mu = m_\tau$ is not the mass of the physical states $\mu$ and $e$, 
which has to be taken as the value(s) at the scale(s) equal to their 
mass(es).  However, at any lower scale, $\mu < m_\tau$, the single 
massive eigenstate $\balpha(\mu)$ of the rotating mass matrix will 
have rotated to a direction different to that of ${\bf v}_\tau$, its 
direction at $\mu = m_\tau$.  It will then no longer be orthogonal to 
the plane spanned by ${\bf v}_\mu$ and ${\bf v}_e$.  But at any scale,
the plane spanned by the two zero eigenvectors is always orthogonal to
the massive eigenvector, so that the state vectors ${\bf v}_\mu$ and
${\bf v}_e$ cannot both remain in the zero eigenspace at this lower
scale. 
Hence, we conclude
that ${\bf v}_\mu$ and ${\bf v}_e$ cannot both be eigenstates with 
zero eigenvalue of the mass matrix at any scale $\mu < m_\tau$, and
confirm the assertion made at the end of the last paragraph.  

At least one of the states ${\bf v}_\mu$ and ${\bf v}_e$ must have a 
nonvanishing component in the direction of the massive state $\balpha(\mu)$ 
for any $\mu < m_\tau$ and acquire thereby a mass, thus contradicting the 
statement above that chiral interations cannot generate nonzero physical 
masses.  That was just a notion gleaned from experience with nonrotating 
mass matrices which is now found inapplicable to rotating mass matrices.  
Notice that since the mass matrix $m$ is still of rank 1, it has at every 
scale 2 linearly independent eigenvectors with eigenvalue zero.  And 
any chiral transformation on these 2 states will leave physics invariant 
so that no chiral property of the mass matrix we started with has ever 
been lost.  Only, by the above analysis, we find that, for the rotating 
mass matrix, in contrast to the nonrotating case, those states on which 
the chiral transformations leave physics invariant are {\it not} the 
physical states, the chiral transformation of which has thus no reason 
to keep the invariance. 

The above example shows that for a rotating mass matrix, the physical 
masses are not in general given just by the eigenvalues of the rotating 
mass matrix, nor the physical states by the eigenvectors, so that zero
eigenvalues do not necessarily imply zero physical masses.  But it begs
the question how the physical masses and states are then to be defined,
to answer which further analysis would be required.  Although such an
analysis has already been given in the literature in the context of
a specific model (DSM) \cite{physcon,ckm,phenodsm,genmixdsm} for fermion 
mixing and mass hierarchy, it pays to review it here outside that context 
so as to exhibit its generality.  Indeed, in doing so, the analysis gains 
also in lucidity, which is a help, for the analysis, though logically 
straightforward in principle, still needs a fair amount of care and 
patience to be carried out.

Let us then go back to the beginning and ask in general terms how 
physical masses and states are to be extracted from a given mass matrix.  
At tree-level, where the concept of a mass matrix originates and where 
the mass matrix is independent of scale, the answer is easy; the measured 
masses are given just by the eigenvalues and the state vectors by the 
corresponding eigenvectors for the various mass states.  On 
renormalization, when the mass matrix depends on scale, however, some 
care is needed, since the eigenvalues and eigenvectors can now be 
scale-dependent, and one needs to specify at what scale(s), if at any, 
these are to be identified as the physical masses and state vectors.
 
Suppose first that the scale-dependent renormalized mass matrix is, 
for some reason, nonrotating, as is the case to a fair approximation 
at least for quarks for the Standard Model, then the answer remains 
relatively simple, since, once diagonalized at some scale, the mass 
matrix will remain diagonal at any other scale, namely of the form: 
$m = {\rm diag}[\lambda_1(\mu), \lambda_2(\mu), \lambda_3(\mu)]$.  Following
then the usual convention that the physical masses are to be measured 
each at the scale equal to the mass itself, we can then identify the 
physical masses $m_i$ of the 3 mass states as respectively the solutions 
to the equations $\lambda_i(\mu) = \mu$.  For leptons, for example, we 
would have $m_\tau = \lambda_1(m_\tau)$, $m_\mu = \lambda_2(m_\mu)$, 
and $m_e = \lambda_3(m_e)$, while the state vectors are given by the 
corresponding eigenvectors which are scale-independent by the initial
nonrotation ansatz.  Notice, however, that even in this case, the 
measured masses are not just the eigenvalues of the same mass matrix but 
of three different matrices representing the mass matrix taken at three 
different scales, and so have departed already from the familiar simple 
notion valid at the tree-level.

What happens next for a rotating mass matrix?  One can still of course 
diagonalize the mass matrix at every scale, but now both the eigenvalues 
$\lambda_i$ and their corresponding eigenvectors, say $\bbeta_i$, will 
depend on scale.  This then raises immediately the question what states 
are to be identified as the physical particle states.  It does not seem 
to make sense to identify just the eigenvectors at some scale as the 
state vectors of the physical particles at that scale, in other words, 
entertaining the concept of scale-dependent physical state vectors.  Take 
again the charged leptons as example.  If we were to identify the physical 
state of the $\tau$ at scale $\mu$, say, as the highest eigenvector
$\bbeta_1(\mu)$, that of the $\mu$ as the next highest $\bbeta_2(\mu)$, 
and that of the $e$ as the last $\bbeta_3(\mu)$, all taken at the same 
scale $\mu$, then since the eigenvectors rotate, what appears as the 
$\tau$ vector at this scale $\mu$ will appear as a mixture of all 3 states 
at a different scale $\mu'$.  In other words, what we thought was the 
$\tau$ at the scale $\mu$ will start decaying into $\mu \gamma$ and 
$e \gamma$ at the other scale $\mu'$, giving thus copious flavour 
violation.  This seems inadmissible.  We ought to give the physical 
states a scale-independent meaning.

In the special case of a nonrotating mass matrix considered above where 
only the eigenvalues but not the eigenvectors depend on scale, it is 
conventional to define, as we did, the masses of the physical particles 
as the eigenvalues taken each respectively at the scale equal to its 
mass.  So one may be tempted similarly to define for the rotating mass 
matrix the state vectors of the physical particles as the respective 
eigenvectors taken each at the scale equal to its mass.  But this also 
will not work.  For if we were to follow this prescription for the 
charged leptons, then one would define the physical $\tau$ state as the 
eigenstate of the mass matrix at scale $\mu = m_\tau$, with the eigenvalue 
$\lambda_1(m_\tau)$ as its mass and the corresponding eigenvector 
$\bbeta_1(m_\tau)$ as its state vector ${\bf v}_\tau$.  Similarly, for 
the $\mu$, we would have $\lambda_2(m_\mu)$ as its mass and the vector 
$\bbeta_2(m_\mu)$ as its state vector ${\bf v}_\mu$.  Now two eigenvectors 
belonging to two different eigenvalues of the same hermitian matrix are 
necessarily orthogonal; hence $\bbeta_1(\mu) \perp \bbeta_2(\mu)$.  But 
the state vector ${\bf v}_\tau = \bbeta_1(m_\tau)$ of $\tau$ has no reason 
to be orthogonal to the state vector ${\bf v}_\mu = \bbeta_2(m_\mu)$ of 
$\mu$, being eigenvectors of the mass matrix $m(\mu)$ taken at different 
values of the scale $\mu$.  Indeed, if the mass matrix rotates, then the 
state vectors so defined for $\tau$ and $\mu$ would not be orthogonal to 
each other, which is physically untenable, since $\tau$ and $\mu$ are 
supposed to be independent quantum states.  It would give rise again to 
unwanted flavour-violations.  

What then has gone wrong?  For the heaviest generation fermion, such as 
$\tau$, the definitions above have no apparent problem; its mass can be 
indeed taken as the highest eigenvalue of the mass matrix and its state 
vector as the corresponding eigenvector, both at the scale $\mu = m_\tau$.  
However, for the next heaviest generation such as $\mu$, a problem begins 
to emerge.  To extract the mass and state vector of $\mu$, the mass 
matrix $m$ has to be taken at an energy scale $\mu < m_\tau$, and at 
these energies, the $\tau$ state becomes unphysical.

To appreciate what this implies, let us recall the familiar parallel case
of the analytic multi-channel $S$-matrix \cite{LeCouteur}, e.g.:
\begin{equation}
S = \left( \begin{array}{ccc} S_{11} & S_{12} & S_{13} \\
                              S_{21} & S_{22} & S_{23} \\
                              S_{31} & S_{32} & S_{33} \end{array} \right).
\label{Smatrix}
\end{equation}
This $3 \times 3$ matrix exists as a mathematical entity at all energies, 
but at energies below the physical threshold of the heaviest channel 1 
where the state 1 becomes unphysical, what represents the physical 
$S$-matrix is just the $2 \times 2$ submatrix at the bottom right corner
labelled by 2 and 3, namely
\begin{equation}
\hat{S} = \left( \begin{array}{cc} S_{22} & S_{23} \\
                                   S_{32} & S_{33} \end{array} \right).
\label{Shat}
\end{equation}
Similarly, at energies below the second heaviest channel 2, the physical 
$S$-matrix is given just by the element $S_{33}$.  In each case, the 
elements of the matrix referring to the higher channels continue to exist 
at the lower energies and represent there just the analytic continuations 
of the physical quantities above the appropriate thresholds but have no 
immediate physical meaning beneath those thresholds.  They cannot, for
example, contribute to unitarity sums, for below those thresholds, the
higher states do not exist as physical states.  

Like the $3 \times 3$ analytic $S$-matrix, the $3 \times 3$ mass matrix 
$m(\mu)$ is a mathematical construct which exists at all energy scales, 
but at energies $\mu$ less than the mass of the heaviest state $m_1$
where this state becomes unphysical, the physical mass matrix is given 
only by the $2 \times 2$ submatrix labelled by the remaining states.  In 
case the mass matrix is nonrotating, or when the rotation is considered 
negligible as in most appliations of the Standard Model, we see that this 
makes no difference to our usual assertions about the physical masses.  
For when the matrix is diagonalized, the truncation of the $3 \times 3$ 
matrix gives for the physical $2 \times 2$ mass matrix for $\mu < m_1$ 
just $\hat{m} = {\rm diag}[\lambda_2(\mu), \lambda_3(\mu)]$.  This will give 
for the physical mass of the second heaviest state $m_2$ again as just the 
solution to the equation $\lambda_2(\mu) = \mu$, as before, and similarly 
also for $m_3$.  

When the mass matrix rotates with changing scale, however, more care is 
needed, for in that case, we recall, we have not yet even identified the 
physical states 2 and 3.  But we do know at least that these states are
independent quantum states to the heaviest state, so their state vectors 
have to be orthogonal to that of the heaviest state.  Hence, it follows
that for a scale less than the mass of the heaviest state, $\mu < m_1$, 
the physical mass matrix, $\hat{m}(\mu)$, has to be the $2 \times 2$ 
submatrix in the 2-dimensional subspace orthogonal to the state vector 
${\bf v}_1$ of the heaviest state.  In particular, we may choose as the 
basis vectors of this orthogonal subspace the vectors $\bbeta_2$ and
$\bbeta_3$ at $\mu = m_1$, which being eigenvectors of the mass
matrix $m$ at $\mu = m_1$ are automatically orthogonal to ${\bf v}_1$.
In this basis, of course, the matrix $\hat{m}(\mu)$ at $\mu = m_1$ is 
diagonal, but because of rotation, it will not remain diagonal at 
lower values of $\mu$.  But $\hat{m}(\mu)$ can be digonalized afresh
at each value of $\mu$ giving eigenvalues, say $\hat{\lambda}_2(\mu),
\hat{\lambda}_3(\mu)$ and their corresponding eigenvectors, say
$\hat{\bbeta}_2(\mu), \hat{\bbeta}_3(\mu)$.  By the same logic as before 
for the heaviest state, we can now define the mass $m_2$ of the second 
heaviest state as the solution to the equation $\hat{\lambda}_2(\mu) 
= \mu$ and the corresponding eigenvector $\hat{\bbeta}_2(m_2)$ as its 
state vector ${\bf v}_2$.  We notice that the vector ${\bf v}_2$ so
defined, being a vector in the orthogonal subspace spanned by the 
chosen basis vectors $\bbeta_2(m_1),\bbeta_3(m_1)$, is automatically 
orthogonal to ${\bf v}_1$, the state vector of the heaviest state.  
In other words, we have now guaranteed that the state vector of $\mu$, 
for example, will be orthogonal to that of $\tau$ and avoided the 
pitfall met with before.  Of course, the identification of ${\bf  v}_2$ 
as the state vector of the second heaviest state also determines the 
state vector ${\bf v}_3$ of the lightest state as the vector orthogonal 
to both ${\bf v}_1$ and ${\bf v}_2$.   

The procedure detailed in the preceding paragraph for identifying the 
masses and state vectors of the physical states applies to any mass
matrix rotating with changing scales.  Let us now specialize to the case 
of a rank 1 mass matrix of the form (\ref{mfact}) above which is of some 
special interest, as will be seen later.  This is easily diagonalized, 
having only one nonzero eigenvalue $\lambda_1(\mu) = m_T(\mu)$ with 
corresponding eigenvector $\bbeta_1(\mu) = \balpha(\mu)$.  The other 
eigenvectors with degenerate eigenvalue zero can be taken as any two 
vectors orthogonal to $\balpha(\mu)$, say $\bbeta_2(\mu),\bbeta_3(\mu)$.  
Following the procedure given above, we then identify the mass of 
the heaviest physical state $m_1$ as the solution to the equation 
$m_T(\mu) = \mu$ and its state vector as $\balpha(m_1)$.  For values of 
$\mu < m_1$, the physical mass matrix according to the above conclusion 
is the truncation of (\ref{mfact}) to the subspace spanned by $\bbeta_2(m_1),
\bbeta_3(m_1)$, i.e.
\begin{equation}
\hat{m}(\mu) = \hat{m}_T(\mu) |\hat{\balpha}(\mu) \rangle 
   \langle \hat{\balpha}(\mu)|,
\label{mhatsp}
\end{equation}
with 
\begin{equation}
\hat{m}_T(\mu) 
   = m_T(\mu) \sqrt{|\langle \balpha(\mu)|\bbeta_2(m_1) \rangle|^2
   + |\langle \balpha(\mu)|\bbeta_3(m_1) \rangle|^2},
\label{mhatT}
\end{equation}
and $\hat{\balpha}(\mu)$ the normalized 2-vector defined as: 
\begin{equation}
|\hat{\balpha}(\mu) \rangle = (m_T(\mu)/\hat{m}_T(\mu)) 
   \left( \begin{array}{c} \langle \balpha(\mu)|\bbeta_2(m_1) \rangle^* \\
   \langle \balpha(\mu)|\bbeta_3(m_1) \rangle^* \end{array} \right).
\label{vhat0}
\end{equation}
This matrix $\hat{m}(\mu)$ vanishes of course when $\mu = m_1$ where the
vectors $\bbeta_2(m_1)$ and $\bbeta_3(m_1)$ are by definition orthogonal
to ${\bf v}_1 = \balpha(m_1)$.  For scales $\mu < m_1$, however, the 
vector $\balpha(\mu)$ would have rotated to another direction giving thus 
a nonzero value to $\hat{m}_T(\mu)$.  The $2 \times 2$ matrix $\hat{m}$ 
which, as we recall, is the physical mass matrix for $\mu < m_1$, is of 
rank 1 as is the original $3 \times 3$ mass matrix $m$.  So the process 
gone through before of identifying mass values and state vectors of physical 
states can be repeated, only now in one less dimension.  One can thus 
immediately conclude that the second heaviest state has a mass $m_2$ given 
by the solution to the equation $\hat{m}_T(\mu) = \mu$, and a state vector 
${\bf v}_2 = \hat{\balpha}(m_2)$.  The process can be repeated again 
to deduce the mass of the lightest state $m_3$.  

The masses $m_2$ and $m_3$ so obtained are seen clearly to have no reason 
to be, and will in general not be, zero when the mass matrix $m$ rotates.
It is thus shown that the two lower generations do naturally acquire 
nonzero masses simply as a result of the rotation of the mass matrix $m$, 
confirming thus the conclusion reached before, only now, as a result of 
the above analysis, one knows exactly what these nonzero masses are
or at least how to compute them, and also the corresponding state vectors.
And in deducing this conclusion, one has nowhere changed the rank 1 
nature of $m$ nor the chiral structure of the action.  At every scale
$\mu$, the mass matrix $m$, being of rank 1, has always 2 eigenstates
with the eigenvalue zero, namely $\bbeta_2(\mu)$ and $\bbeta_3(\mu)$
in the above notation, and a chiral transformation on these 2 states
will leave the action invariant.  Only these states, $\bbeta_2(\mu)$ 
and $\bbeta_3(\mu)$, on which a chiral transformation leaves physics 
invariant, are not to be identified with the physical states which
are the states ${\bf v}_2$ and ${\bf v}_3$.  These physical states 
${\bf v}_2$ and ${\bf v}_3$ are linear combinations of the ``chiral'' 
massless states $\bbeta_2(\mu)$ and $\bbeta_3(\mu)$ at any scale $\mu$ 
but they contain in addition an admixture of the massive eigenstate 
$\bbeta_1(\mu)$ at that scale.  It is this admixture of the massive
eigenstate in the physical states ${\bf v}_2$ and ${\bf v}_3$ which 
gives them each a mass and destroys at the same time the invariance 
under a chiral transformation on them.  Their nonzero masses arise 
purely from the rotation, as the energy scale $\mu$ changes, of the 
vector $\balpha(\mu)$ away from its direction at $\mu = m_1$, i.e. 
the direction of the state vector ${\bf v}_1$ for the heaviest state.  
It thus appears that simply by virtue of the rotation, the mass of the 
heaviest state has leaked a little into the lower generations to give 
them small but nonvanishing masses, hence the rather fanciful name of 
``leakage mechanism'' we have coined earlier for it \cite{ckm}.     

We notice that to actually evaluate the physical masses arising from 
a rotating mass matrix in accordance to the above procedure, we shall
need to know the rotation matrix $U(\mu, \mu_0)$ as defined in
(\ref{rotation}) above, which will depend on the underlying theory.
Besides, we need also to specify at exactly what scales the physical
masses are to be measured, which in the above analysis we have chosen,
following the usual convention, to be the scales equal to the masses
themselves.  This last is reasonable if we are dealing with freely
or quasi-freely propagating particles like the charged leptons or the
heavy quarks, but may not be the most convenient for confined objects
like the light quarks $u$ and $d$.  However, for the general assertion 
that nonzero masses can result from a chiral invariant but rotating 
mass matrix, we shall not need to be specific about either the rotation
matrix $U(\mu, \mu_0))$ nor the scales at which the physical masses are
measured.  The assertion would follow so long as the mass matrix does 
rotate and the physical masses of the different physical states are to 
be measured at different scales, as indicated already in the remarks 
made right at the beginning. 

That nonzero masses can result from a rotating chiral invariant mass 
matrix is not just a theoretical curiosity but can lead to some quite
interesting physical consequences.  The reason is that there are many
instances when from various theoretical considerations it may seem 
physically desirable to start with a chiral invariant mass matrix for
fermions but is hindered by the experimental fact that the fermions we 
see seem all to possess nonzero masses, although their masses are in
some cases rather small.  However, it would now appear from above that 
if we allow the mass matrix to rotate then we can both keep the chiral 
invariance and have nonzero masses for our physical particles, thus 
bypassing the apparent contradiction.  

As an example of such instances where a chiral invariant mass matrix
seems desirable, consider first the strong CP problem.  This arises 
from the fact that colour gauge invariance in QCD admits in principle 
a CP-violating term in the action of the form;
\begin{equation}
{\cal L}_\theta = -\frac{\theta}{64 \pi^2} \tilde{F} F,
\label{Ltheta}
\end{equation}
associated with topologically non-trivial field configurations, where 
$\theta$ is a real but otherwise arbitrary parameter \cite{WeinbergII}.  
Such a term in the action would exhibit itself, for example, in a 
nonvanishing electric dipole moment of the neutron estimated to be of 
the order \cite{WeinbergII}:
\begin{equation}
d_n \sim |\theta| e m_\pi^2/m_N^3 \sim 10^{-16} |\theta| \  {\rm e \  cm}.
\label{edm}
\end{equation}
The present experimental limit for $d_n$ has already been pushed down to
less than $3 \times 10^{-26}$ e cm, which means that this free parameter 
$\theta$ in the theory, if it really exists, will have to be assigned a value:
$|\theta| < 10^{-10}$.  It would appear therefore that nature has some 
hidden mechanism for suppressing this angle $\theta$ which has not yet 
been accounted for in the standard formulation of chromodynamics.  The 
favourite mechanism suggested is to supplement colour symmetry by an
additional $U(1)$ symmetry \cite{PecceiQ}, the breaking of which, however, 
would give rise to a new particle called the axion \cite{PecceiQ,Weinbergb}, 
which has been searched for experimentally but never yet observed.  Besides,
new experiments are being carried out which are expected to push down
further the limit of the electric dipole moment of the neutron, which
is already making it uncomfortable for most current suggestions for the
suppression mechanism \cite{KGreen}.  It would thus be of interest to 
explore other possibilities for suppressing the $\theta$ angle or for
eliminating it altogether.

The reason why a rotating fermion mass matrix matters in the strong CP 
problem is that the $\theta$ angle term (\ref{Ltheta}) in the action can 
be removed by a chiral transformation on the fermion fields which leaves 
the physics invariant provided that the quark mass matrix has at least 
one zero eigenvalue \cite{WeinbergII}.  Unfortunately, as far as known, 
no quark can be assigned a zero mass in the current interpretation of 
the existing experimental data, and would thus be at variance with the 
above proviso, if chiral invariance necessarily implies a zero mass for
a physical state.  However, in the above analysis, one sees that, if the 
mass matrix rotates, then the physically measured masses of all physical 
fermion states can be nonzero even though the mass matrix appearing in 
the action has zero eigenvalues.  That being the case, it would seem to 
offer a possible resolution to the above dilemma, at least in principle
\cite{Jakov}.

As a second example, consider the chiral symmetry breaking problem in QCD.  
As noted already, the QCD action is ``almost'' chiral invariant.  Indeed, 
if the light quarks $u$ and $d$ actually have zero mass, then the QCD 
action would be invariant under chiral $su(2) \times su(2)$.  It thus
seem attractive to entertain the notion that the QCD action may in fact 
be invariant under chiral $su(2) \times su(2)$ to start with, but then 
undergoes a spontaneous breaking of this symmetry to give the $u$ and
$d$ quarks each a mass.  Although the actual mechanism for the spontaneous
breaking of this chiral $su(2) \times su(2)$ has not been fully understood,
the idea has generated a host of important results  too numerous to be
here enumerated \cite{WeinbergII}.  Now, if it were true, as suggested by
the above analysis, that the light quarks $u$ and $d$ can acquire each a
physical mass different from zero without the action ever losing its chiral 
invariance so long as the mass matrix rotates, would it not then cast a new
light on to the problem?  The masses of $u$ and $d$ arising from rotation
via the ``leakage mechanism'' would be naturally small.  In other words, 
their smallness would appear as a consequence of chiral invariance instead 
of being a hindrance to it.

Of course, this new angle for looking at the strong CP and chiral symmetry 
breaking problems would be no more than exchanging these two mysteries for
another, namely that of mass matrix rotation, unless one can find a viable
theoretical reason why the fermion mass matrix should rotate, or else some
evidence in nature that it does do so.  It turns out that both such exist 
though both are as yet of a circumstantial nature.  Nevertheless, they 
seem to us already to be a sufficient incentive for the rotation scenario 
to be seriously entertained.

These arise as follows.  Since, according to the above analysis, the lower
generations can acquire each a mass by ``leakage'' from the generation 
above, it follows that only the heaviest generation needs be given a mass 
(i.e. starting with a mass matrix $m$ of rank 1) for all generations to 
end up with nonzero physical masses.  Now given the empirical fact that
fermions of the heaviest generation are in every known case very much 
heavier than the others, a rank 1 mass matrix has long been taken as a 
good starting point for a phenomenological description \cite{Fritzsch}. 
The ``leakage mechanism'' from rotation now provides one with a concrete 
procedure for actually producing finite masses for the lower generations 
starting from a rank 1 mass matrix.  Such a scenario is phenomenologically 
particularly attractive for the following reason.  Since the masses of 
each lower generation arise only as consequences of ``leakage'' from those
of the generation above, they are expected to have progressively smaller 
values, dropping by large factors from generation to generation.  In other 
words, we have here an immediate qualitative explanation for the fermion 
mass hierarchy observed in experiment.  Furthermore, since state vectors 
for different flavours are to be defined each at the scale equal to its
mass, it follows that the state vectors of up-states will not be aligned 
to those of down-states, given their different masses, even if their mass 
matrices are always aligned at the same scale.  For example, the state 
vector ${\bf v}_t$ of the $t$-quark is the first eigenvector of the mass 
matrix of $U$-type quarks evaluated at $\mu = m_t$, while the state vector 
${\bf v}_b$ of the $b$-quark would be the first eigenvector of the mass 
matrix of $D$-type quarks but evaluated at $\mu = m_b$.  Hence, even if 
the mass matrices of the 2 quark types are always aligned at the same 
$\mu$, the 2 state vectors ${\bf v}_t$ for $t$ and ${\bf v}_b$ for $b$ 
will not be aligned, meaning that there will be nontrivial mixing between 
the $t$ and $b$ states.  In other words, a single rotating rank 1 mass 
matrix has already the potential to explain not only the fermion mass 
hierarchy experimentally observed but also the intriguing mixing pattern 
between the $U$ and $D$ flavours.

The idea outlined in the preceding paragraph for explaining the fermion
mass hierarchy and mixing pattern can be put to empirical test in two 
ways.  First, starting with the experimental quark and lepton masses 
and mixing angles, and interpreting them as arising from a single 
rotating rank 1 mass matrix, one asks whether the result is consistent 
with all the data points lying on a smooth rotation curve.  This was done 
and the answer is affirmative within experiemtnal errors \cite{cevidsm}.  
Conversely, one can start by constructing a model for rotation giving a 
rotating rank 1 mass matrix depending on some paramters, and then proceed 
to fit the experimental data with the model.  This was done with a model 
called the Dualized Standard Model (DSM) which was able to give a good 
fit to nearly all the mass ratios and mixing angles with only 3 adjustable 
real parameters \cite{phenodsm,genmixdsm}.  We find these tests rather 
compelling, given that the fermion mass hierarchy and mixing pattern have 
otherwise no generally accepted explanation, and can, we think, be taken 
as at least circumstantial evidence for mass matrix rotation.  

As to theoretical justification for why the fermion mass matrix should 
rotate and at such speed as to produce the above phenomena, our judgment
is bound to be a little subjective, given our past experience.  We can
say, however, that the model DSM cited above for fitting data was meant 
to be only phenomenological, having been constructed with the object in 
mind, and thus contains some ad hoc assumptions while satisfying no strict 
demand for internal consistency.  Besides, it is seen \cite{phenodsm}
that its apparent success as outlined above does not depend so much on 
its details but largely just on the fact that the rotating fermion mass 
matrix it produced has 2 rotational fixed points, one at $\mu = 0$ and 
the other at $\mu = \infty$.  However, a new self-consistent model has
now been constructed on a firmer theoretical basis \cite{prepsm}.  It
has as its motivation an explanation of some of the Standard Model's
basic features, and incorporates 't~Hooft's confinement picture for 
symmetry-breaking \cite{tHooft,Bankovici} while purporting to give a 
new geometrical meaning to Higgs fields.  It has thus a very different 
structure from the previous model.  Nevertheless, the new model leads 
logically also to a rotating fermion mass matrix of rank 1 with still 
the desired fixed points at $\mu = 0, \infty$.  Besides, it has overcome 
some of DSM's shortcomings and gained some new good features such as the 
possibility of a CP-violating phase.  It is thus hopeful that the fit to 
experimental data now being carried out may equal or perhaps even surpass 
that obtained before with the DSM.  If this results, then the empirical 
observations made in the preceding paragraphs would have been put on a 
firmer theoretical footing. 

Although the evidence for mass matrix rotation as outlined above, whether
empirical or theoretical, is as yet only circumstantial, it appears to us 
already sufficient to suggest that this possibility be taken seriously.  
That being the case, it may in turn cast a new light on to the strong CP 
and chiral symmetry problems, as observed above.

\end{document}